\documentclass[pra,reprint,showpacs,twocolumn,superscriptaddress,floatfix]{revtex4-1}

\usepackage{epsfig}
\usepackage{amsmath}
\usepackage{amssymb}
\usepackage{amsfonts}
\usepackage{mathptmx}
\usepackage{dcolumn}
\usepackage{eucal}
\usepackage{bm}
\usepackage{color}
\usepackage[colorlinks,linkcolor=blue,citecolor=blue]{hyperref}
\usepackage{epstopdf}
\usepackage{bbold}
\usepackage{url}
\usepackage{mathtools}
\usepackage{physics}
\usepackage{appendix}
\usepackage{dsfont}

\def \ket#1{\mathinner{|{#1}\rangle}}
\def \bra#1{\mathinner{\langle{#1}|}}

\newcommand{\Prob}{\mathcal{P}}

\newcommand{\Glambda}{\mathcal{G}_\lambda}

\begin{document}

\title{Quantum simulations of macrorealism violation via the QNDM protocol}

\author{D. Melegari}
\affiliation{Dipartimento di Fisica, Universit\`a di Genova, via Dodecaneso 33, I-16146, Genova, Italy}
\affiliation{INFN - Sezione di Genova, via Dodecaneso 33, I-16146, Genova, Italy}
\author{M. Cardi}
\affiliation{Dipartimento di Fisica, Universit\`a di Genova, via Dodecaneso 33, I-16146, Genova, Italy}
\affiliation{INFN - Sezione di Genova, via Dodecaneso 33, I-16146, Genova, Italy}
\author{P. Solinas}
\affiliation{Dipartimento di Fisica, Universit\`a di Genova, via Dodecaneso 33, I-16146, Genova, Italy}
\affiliation{INFN - Sezione di Genova, via Dodecaneso 33, I-16146, Genova, Italy}

\date{\today}

\begin{abstract}
The Leggett-Garg inequalities have been proposed to identify the quantum behaviour of a system; specifically, the violation of macrorealism.
They are usually implemented by performing two sequential measurements on quantum systems, calculating the correlators of such measurements and then combining them arriving at Leggett-Garg inequalities.
However, this approach only provides sufficient conditions for the violation of macrorealism.
Recently, it was proposed an alternative approach that uses non-demolition measurements and gives both a necessary and sufficient condition for the violation of macrorealism.
By storing the information in a quantum detector, it is possible to construct a quasi-probability distribution whose negative regions unequivocally identify the quantum behaviour of the system.
Here, we perform a detailed comparison between these two approaches.
The use of the IBM quantum simulators allows us to evaluate the performance in real-case situations and to include both the statistical and environmental noise.
We find that the non-demolition approach is not only able to always identify the quantum features but it requires fewer resources than the standard Leggett-Garg inequalities.
In addition, while the efficiency of the latter is strongly affected by the presence of the noise, the non-demolition approach results incredibly robust and its efficiency remains unchanged by the noise.
These results make the non-demolition approach a viable alternative to the Leggett-Garg inequalities to identify the violation of macrorealism.
\end{abstract}

\maketitle

%%%%%%%%%%%%%%%%%%%%%%%%%%%%%%%%%%%%

\section{Introduction}

The ability to distinguish between the quantum and classical behaviour of a system is of paramount interest both in the foundations of quantum mechanics and in quantum technologies.
From a fundamental perspective, we aim to identify which features are intrinsically quantum and how a quantum system eventually becomes classical, e.g., due to interaction with an environment.
In the context of quantum technologies, we are interested in exploiting such features to enhance computation or information exchange.
In this regard, we need a formal and quantitative method to identify the quantumness of a system.
Two key tools have emerged in this direction: Bell's inequalities \cite{Bell1964} and the Leggett-Garg inequalities (LGIs) \cite{Leggett1985}.
The former demonstrates that quantum mechanics is a nonlocal theory as a consequence of entanglement.
The latter identifies violations of macrorealism (MR) due to the presence of coherent superpositions of quantum states \cite{maroney2014,Schmid2024}.
The MR condition consists of a set of requirements that classical systems satisfy but quantum ones do not.
Therefore, a violation of the LGIs allows us to highlight the quantum behaviour of a system.
Despite their significance, LGIs provide only a sufficient condition for the violation of the MR condition.
In other words, if the LGIs are violated, we can confidently assert that the system exhibits quantum properties.
However, if they are satisfied, we cannot necessarily rule out quantum behaviour.

For this reason, significant efforts have been made to identify the necessary conditions for violating the MR condition \cite{Halliwell2019, Halliwell2016, Clemente2015, Clemente2016}.
Recently, a clear, efficient and operational approach has been proposed that provides both a necessary and sufficient condition \cite{solinas2024}.
This protocol is based on quantum non-demolition measurements (QNDM)~\cite{Braginsky1980, BraginskyBook, Caves1980, CavesRevModPhys} and was introduced in Refs.~\cite{solinas2013work,solinas2015fulldistribution,solinas2016probing,solinas2021,solinas2022,GherardiniTutorial}.
The main advantage of the QNDM protocol lies in its experimental feasibility.
Since it was designed as a sequence of concrete experimental steps, its operational meaning is clearer compared to other more abstract approaches \cite{Clemente2015, Clemente2016, Halliwell2016}.

The protocol's outcome is a quasi-probability distribution (QPD) density, which can exhibit negative regions.
As with the well-known Wigner function, the presence of negative regions in the QPD signals the presence of quantum effects.
While this connection was already known~\cite{solinas2013work,solinas2015fulldistribution,solinas2016probing,solinas2021,solinas2022,GherardiniTutorial}, recent studies have shown that the presence of negative regions provides both a necessary and sufficient condition for the violation of the MR condition \cite{solinas2024}.
This result bridges the gap in LGIs, as it allows us to state that the absence of negative regions implies that MR is satisfied and the system exhibits classical properties.

In this article, we further develop the ideas presented in Ref. \cite{solinas2024}.
First, we consider a realistic scenario in which experiments are repeated a finite number of times.
This introduces statistical uncertainty into the results, which ultimately affects the ability to identify violations of macrorealism.
Through detailed numerical simulations using IBM’s \texttt{Qiskit} language \cite{IBM_docs}, we compare the efficiency of the QNDM approach and LGIs in terms of the resources required.
We find that the QNDM approach is not only more effective in detecting quantum effects (as it provides a necessary and sufficient condition), but it also requires fewer resources to execute and is more robust against statistical noise.

The reason behind this surprising result lies in how the QPD is computed using a Fourier transform applied to the experimental data.
The Fourier transform proves to be highly robust against imprecisions in the experimental data, such as statistical noise arising from repeated measurements.
This robustness allows for a reduction in the number of experimental repetitions without compromising the critical features of the QPD—specifically, the presence of negative regions.

We also extend our comparison to scenarios where environmental noise is present.
To this end, we take advantage of IBM’s \texttt{Qiskit} simulations \cite{IBM_docs}, which allow for the inclusion of realistic environmental noise.
The results further confirm the superiority of the QNDM protocol over LGIs in terms of both efficiency, robustness and resource requirements.

The paper is structured as follows.
In Sec. \ref{sec: MR_conditions}, we introduce the MR condition and discuss its physical significance.
Section \ref{sec: MR_violation} provides an overview of how LGIs and the QNDM protocol detect MR violations.
In Sec. \ref{sec: num_impl}, we present numerical simulations for both protocols, with and without environmental noise, and offer a detailed comparison in terms of resource consumption.
Finally, in Sec. \ref{sec: conclusions}, we summarize our results and draw conclusions.

%%%%%%%%%%%%%%%%%%%%%%%%%%%%%%%%%%%

\section{Macrorealism}
\label{sec: MR_conditions}
The concept of macrorealism was introduced in $1985$ by Leggett and Garg to distinguish the features of a classical system from those of a quantum system \cite{Leggett1985}.  
Every macrorealistic system should satisfy three conditions:  
$1)$ Macrorealism {\it per se} (MRps), which states that at any given time, a macroscopic object with two available states is in a definite one of those states.;
$2)$ Non-Invasive Measurability (NIM), which states that it is possible, in principle, to determine in which state the system is, without any effect on the state and the subsequent dynamics;
$3)$ Induction or "arrow of time", which states that future measurements cannot affect the present state \cite{Halliwell2019}.  

Usually, when the measurements performed on the system are sequential, the NIM condition is substituted by the No Signaling In Time (NSIT) conditions \cite{Clemente2015,Clemente2016, Kofler2013}.  
The NSIT condition can be seen as a statistical version of NIM \cite{Halliwell2017}.  

The MRps was later reformulated more clearly, stating that it is satisfied if there cannot be \textit{coherent superpositions} and the state of the system is always diagonal in the basis that diagonalizes the observable to be measured~\cite{Schmid2024, Knee2016}.  

Under these three assumptions, a set of Leggett-Garg inequalities (LGIs) can be derived \cite{Leggett1985, Emary_2014}.
They are formally similar to the Bell-Clauser–Horne–Shimony–Holt (CHSH) inequality are satisfied by classical systems but might be violated by quantum systems \cite{Bell1964, CHSH}.
From their proposal, the LGIs have become the privilege tool to identify the presence of coherent superposition in quantum system.

%%%%%%%%%%%%%%%%%%%%%%%%%%%%%%%%%%%%%%

\section{Violation of macrorealism: Leggett-Garg inequalities and non-demolition protocols}
\label{sec: MR_violation}

The physical situation we analyse is the following.
Suppose we have a two-level quantum system $S$ evolving under a unitary transformation in a time interval $0 \leq t \leq T$. At times $t_0\leq t_1 \leq t_2=T $, we measure a generic observable $\hat{A}$ (Hermitian operator) which has eigenvalues $a_i = \pm 1$ with the corresponding eigenstates denoted with $\ket{i}$, i.e., , $\hat{A}\ket{i} = a_i\ket{i}$.
In between two measurements, the system evolves with a unitary operator $\hat{U}_i$ with $i=1, 2$.
In general, we have that $[\hat{A}, \hat{U}_i] \neq 0$ so that there is a complex interplay between the measurement and the evolution.
Notice that this scheme accounts also for the situation in which we measure a sequence of non-commuting observables $\hat{A}$, $\hat{B}$ and $\hat{C}$ without evolution in between the measurements.

With this set-up, we want to identify the presence of quantum behaviour of the system by using a experimantal feasible protocol.

%%%%%%%%%%%%%%%%%%%%%%%%%%%%%%%%%%%%%%%%%%%%%%%%%%%%

\subsection{Leggett-Garg inequalities}
\label{sec: LGIs}

As a first method to detect the violation of macrorealism, we consider the LGIs.
They are constructed considering quantum correlators $C_{ij}$ between the outcomes of the projective measurements at different times $t_i$ and $t_j$.

Using the Heisenberg picture, we denote as $\hat{A}(t_i) = \hat{U}_i^\dagger \hat{A} \hat{U}_i$ so that the correlator can be written as:
\begin{align}
C_{ij} = \text{Tr}[(\hat{A}(t_i)\hat{A}(t_j) + \hat{A}(t_j)\hat{A}(t_i))\hat{\rho}] = \sum_{i,j} a_i a_j P(a_i, a_j),
\end{align}
where \( P(a_i, a_j) \) is the probability of observing the eigenvalue \( a_i \) at time \( t_i \) and \( a_j \) at time \( t_j \).

With the three possible correlators, we can calculate the Leggett-Garg parameter $K_3$ as:
\begin{align}
\label{eq: LGI_3}
K_3 = C_{01} + C_{12} - C_{02}.
\end{align}
For a classical system, we necessarily have $-3 \leq K_3 \leq 1$ \cite{Emary_2014} so that a violation of this inequality indicates a behaviour inconsistent with classical macrorealism.

Alternatively, we can choose two other equivalent combinations of the correlations, which are likewise bounded from $-3$ to $1$ \cite{Emary_2014}. They are:
\begin{align}
 K_1 &=  C_{12} - C_{01} + C_{02} \nonumber \\
 K_2 &= -C_{12} - C_{01} - C_{02}.
\end{align}

%%%%%%%%%%%%%%%%%%%%%%%%%%%%%%%%%%%%%%%%%%%%%%%%%%%%

\subsection{QNDM protocol}
\label{sec: qndm_prot}

An alternative way to extract information about the values of $\hat{A}$ are different times is through a Quantum Non-Demolition Measurements (QNDM) \cite{Bednorz2012}.
Here, the fundamental idea is to store the information about the observable $\hat{A}$ in a phase of a quantum detector without perturbing the expected system evolution.
The final measurement of the detector phase gives information about the sequential measurement of $\hat{A}$ at different times.

Let $\hat{p}$ be an operator acting on the degrees of freedom of the detector and $p$ and $\ket{p}$ its eigenvalues and eigenstates, respectively, $\hat{p} \ket{p} = p \ket{p}$.
From the QNDM perspective, the process of measurement at three times $t_0,t_1$ and $t_2$ is equivalent to consider a sequence of interaction between the system and the detector $\hat{u}(\lambda) = \exp \{i (\lambda/2) \hat{A} \otimes \hat{p}\} $ - $\lambda$ is the coupling constant between the system and the detector - followed by the evolution of the system $\hat{U_i} $. Thus, the total operator acting on the system$+$detector is given by:
\begin{equation}\label{eq:tot_evo}
 U_{\text{tot}} = \hat{u} \hat{U}_2 \hat{u} \hat{U}_1 \hat{u} = e^{i\frac{\lambda}{2} \hat{A} \otimes \hat{p}}  \hat{U}_2 e^{i\frac{\lambda}{2} \hat{A} \otimes \hat{p}}  \hat{U}_1 e^{i\frac{\lambda}{2} \hat{A} \otimes \hat{p}}. 
\end{equation}
Let's denote $\ket{\psi^S_0} = \sum_{i}\psi^0_i\ket{i}$ the initial state of the system and $\ket{\psi_{0}^{D}}$ the initial state of the detector. 
For our purposes, we will take $\ket{\psi_{0}^{D}}$ as an equal superposition of $N$ eigenstates of $\hat{p}$. 
The initial state of the system$+$detector is taken to be the product state $\ket{\Psi_0} =\ket{\psi^S_0}\otimes\ket{\psi^D_0}$.

The final state of the composed system and detector under the evolution in Eq. \eqref{eq:tot_evo} reads \cite{solinas2024}
\begin{equation}
    \ket{\Psi} = \dfrac{1}{\sqrt{N}}\sum_{p}\sum_{ijk}e^{i\frac{\lambda p}{2}(a_i+a_j+a_k)}U_{kj}U_{ji}\psi^0_i\ket{i}\ket{p}.
\end{equation}

The phase accumulated between two states $\ket{\pm}$ of the detector $\Glambda$ is 
\begin{align}\label{eq: Glambda formal}
 \Glambda  = \frac{\Tr\qty[\qty(\mathbb{I} \otimes \bra{p} )  \hat{R} \qty(\mathbb{I} \otimes \ket{-p})]}{\mel{p}{\hat{r}_0}{-p}}  
\end{align}
where $\hat{R} = \ketbra{\Psi}{\Psi}$, and $\hat{r}_{0} = \ketbra{\psi_{0}^{D}}{\psi_{0}^{D}}$ are the final total and the detector initial density matrix, respectively.
The function $\Glambda$ is a quasi-characteristic function of the measurement of $\hat{A}$ in time \cite{solinas2015fulldistribution,solinas2016probing,solinas2013work, solinas2024}.
Therefore, by calculating the inverse Fourier transform of $\Glambda$, we obtain the QNDM quasi-probability distribution 
\begin{equation}\label{eq: P_ND}
\begin{split}
 \mathcal{P}_{\text{ND}}(\Delta) &= \frac{1}{2\pi} \int e^{-i\lambda \Delta}  \Glambda \dd{\lambda} \\
    &= \sum_{ijklm} P_{ND}(i,j,k,l,m)\delta(\Delta - \Delta_{ijklm})
\end{split}
\end{equation}
where $\Delta_{ijklm} = a_k +(a_i+a_j+a_l+a_m)/2$ and $P_{ND}$ are the weights of the distribution. 

The $ \mathcal{P}_{\text{ND}}(\Delta)$ is a quasi-probability density distribution because it can have negative regions.
As in the case of the Wigner function, these negative regions highlight the presence of quantum phenomena.
More specifically, they are directly linked to the presence of coherent superpositions during the dynamics \cite{solinas2015fulldistribution,solinas2016probing,solinas2024}.
As a consequence, the presence of negative regions signals the violation of MRps.

It must be noticed that, as discussed in in Ref. \cite{solinas2024}, by construction the quasi-probability $\mathcal{P}_{\text{ND}}$ satisfy the NSIT condition. Therefore, any violation of MRps leads to the violation of MR.

%%%%%%%%%%%%%%%%%%%%%%%%%%%%%%%%%%%%%%%%%%%%%

\section{Numerical study}
\label{sec: num_impl}

To compare the efficiency of the LGI and QNDM to identify the violation of macroreaslim, we consider a simple example already discussed in Refs. \cite{Halliwell2019, solinas2024, sissaleggettgarg}.

Denoting with $\hat{\sigma}_i$ with $i=x, y, z$ that Pauli operators, we consider $\hat{A} = \hat{\sigma}_{z}^{D}$ the dynamics generated by the Hamiltonian $\hat{H} = \omega\hat{\sigma}_x^S/2$ (with $\hbar=1$).
The measurements occur at times $t_0=0$, $t_1=\tau$ and $t_2=2\tau$ and we choose the initial state as:
\begin{equation}
 \ket{\psi^{S}_{0}} = \frac{\ket{0} + i \ket{1}}{\sqrt{2}},
\end{equation}
where $\ket{0}$ and $\ket{1}$ are eigenstate of $\hat{\sigma}_z$.

%%%%%%%%%%%%%%%%%%%%%%%%%%%%%%%%%%%%%%%%%%%%%%%%

\subsection{Leggett-Garg numerical implementation}

Given the definitions above, the correlations $C_{ij}$ take the form  
\begin{equation}
    C_{ij} = \cos\qty(\omega (t_{i} - t_{j}))
\end{equation}  
as reported in Refs. \cite{ Halliwell2016, solinas2024, sissaleggettgarg}. Consequently, the LG parameter $K_3$ is given by:  
\begin{equation}\label{eq: K3 analytical}
    K_3 = C_{01} + C_{12} - C_{02} =  2\cos\qty(\omega \tau) - \cos\qty(2\omega \tau).
\end{equation}  
We study the behaviour of $K_3$ as a function of $\omega \tau$
Any value falling outside the range  $-3 \leq K_3 \leq 1$ implies a violation of the LGIs.  

\begin{figure}
    \begin{center}
        \includegraphics[width=0.8\linewidth]{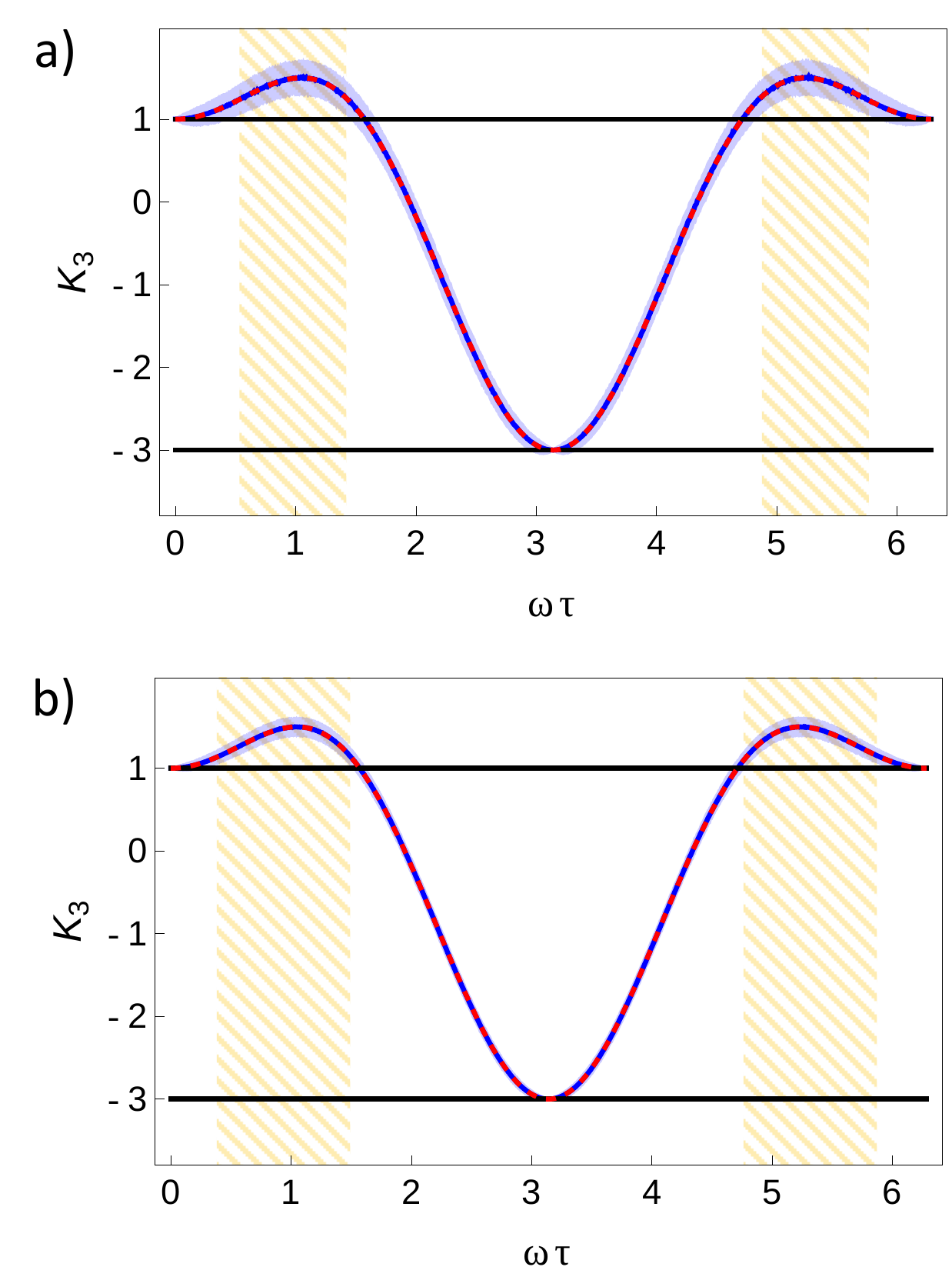}
    \end{center}
    \caption{
        Simulation results for the LG parameter $ K_3 $ in a noiseless scenario, using $ N_{\text{shots}} = 10^3 $ (a) and $ N_{\text{shots}} = 10^4 $ (b). The solid blue line represents the average simulated curve obtained over $N = 100$ repetitions, while the dashed red line corresponds to the theoretical prediction from Eq.~\eqref{eq: K3 analytical}. The blue-shaded region indicates statistical uncertainty, while the yellow-shaded region highlights the range of $ \omega \tau $ where, considering statistical errors, a violation of the LGIs is confidently detected.
    } 
    \label{fig: K3_fig}
\end{figure}

To investigate this phenomenon, we performed quantum simulations using IBM’s quantum computing framework, \texttt{Qiskit} \cite{IBM_docs}. The implementation details of the LG simulations can be found in Appendix \ref{sec: qiskit_implementation} and the results of these simulations are depicted in Fig.~\ref{fig: K3_fig}.

To obtain the simulated curves, we performed $N = 100$ independent runs for each value of $\omega \tau$ and computed the average of the resulting $K_3$ values (solid blue line). In contrast, the dashed red line represents the theoretical curve given by Eq.~\eqref{eq: K3 analytical}. In absence of environmental noise, the two curves overlap.

A key aspect of our study is determining whether the violation of LGIs can be confidently detected within a single experimental run of $N_{\text{shots}}^{\text{LG}}$ measurements. The impact of statistical uncertainty is illustrated by the blue-shaded regions in Fig.~\ref{fig: K3_fig}. Specifically, Fig.~\ref{fig: K3_fig}a) shows results for $N_{\text{shots}}^{\text{LG}} = 10^3$, while Fig.~\ref{fig: K3_fig}b) corresponds to $N_{\text{shots}}^{\text{LG}} = 10^4$.  
To quantify this effect, the yellow-shaded region highlights the values of $\omega \tau$ where LGI violations are clearly distinguishable despite statistical errors. As expected, increasing the number of shots reduces statistical fluctuations, leading to better agreement between simulations and theory.  

In an ideal scenario with zero statistical noise, LGI violations are expected in the ranges $0 \leq \omega \tau \leq \pi/2$ and $3\pi/2 \leq \omega \tau \leq 2\pi$, covering $50\%$ of the total $\omega \tau$ range. However, in realistic conditions where $N_{\text{shots}}^{\text{LG}}$ is finite, statistical uncertainty reduces the detectable violation region. For $N_{\text{shots}}^{\text{LG}} = 10^3$, the violation region covers approximately $32\%$ of the total range, whereas for $N_{\text{shots}}^{\text{LG}} = 10^4$, it extends to $38\%$.  

%%%%%%%%%%%%%%%%%%%%%%%%%%%%%%%%%%%%%%%%%%%%%%%%%

\subsection{QNDM numerical implementation}\label{sec: QNDM_numerical}

For the implementation of the QNDM protocol, we use a two-level quantum detector.
We take  $\hat{p} = \hat{\sigma}_z^D$, the initial state to be $ \ket{\psi_{0}^{D}} = (\ket{0^{D}} + \ket{1^{D}})\sqrt{2}$ and the interaction operator as:
\begin{equation}
\hat{u}(\lambda) = \exp\{i(\lambda/2) \hat{\sigma}_z^S \otimes \hat{\sigma}_z^D\}.
\end{equation}
Its implementation in \texttt{Qiskit} is provided in Appendix \ref{sec: qiskit_implementation}.

\begin{figure}
    \begin{center}
    \includegraphics[width=0.9\linewidth]{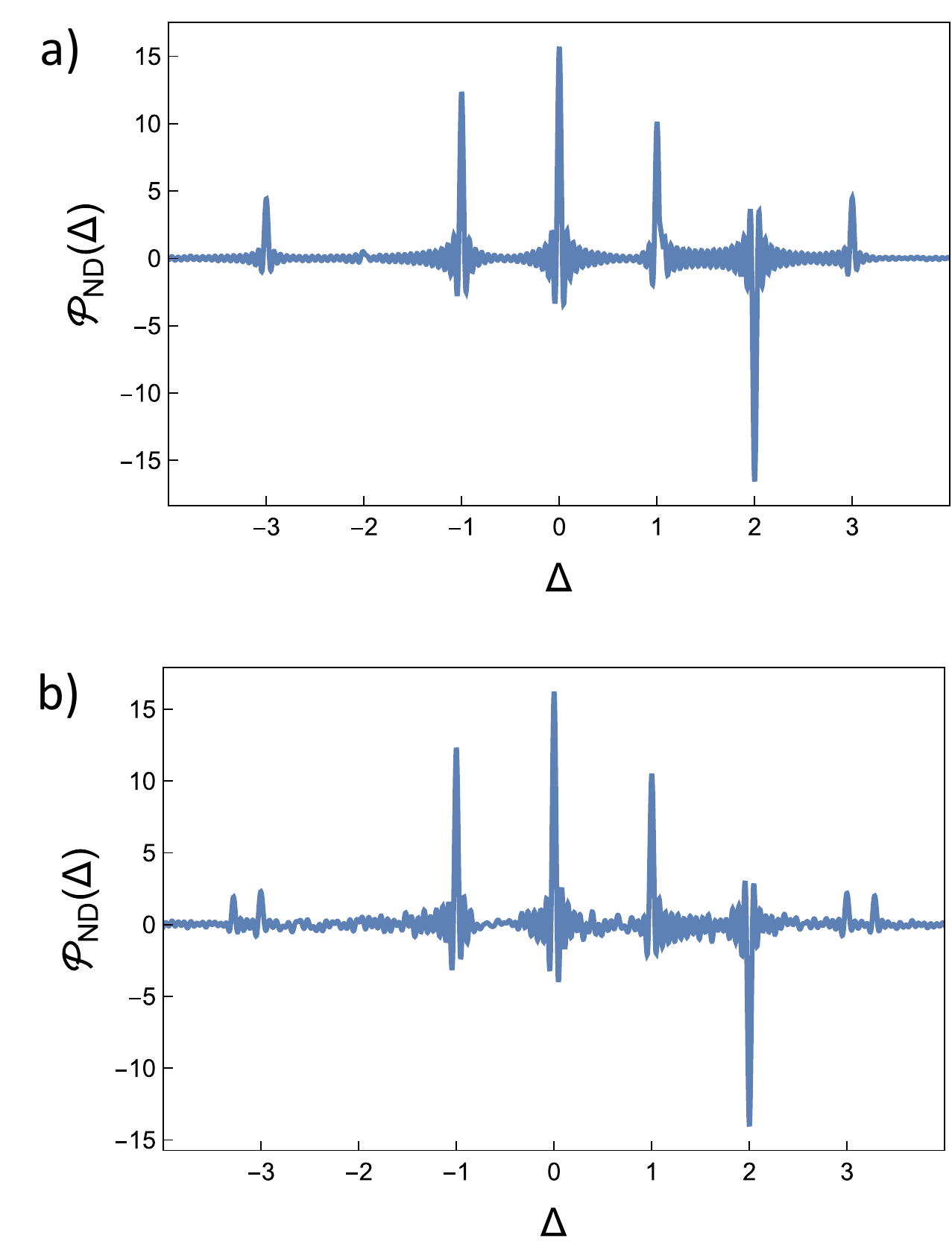}
       \end{center}
    \caption{
Plot of the quasi probability $\Prob_{\text{ND}}(\Delta)$ for $\Delta \lambda = 0.1 $ and $N_{\text{shots}} = 10^{3}$ (a) and for $\Delta \lambda = 1$ and $N_{\text{shots}} = 10^2$ (b).} 
    \label{fig: QPDV1}
\end{figure}

The advantage of the QNDM is that the presence of negative regions in the $\Prob_{\text{ND}}(\Delta)$, independently from their amplitude or details, is enough to identify the violation of macrorealism.
This gives an intrinsic robustness to the protocol which is confirmed in Fig. \ref{fig: QPDV1}.

%%%%%%%%%%%%%%%%%%%%%%%%%%%%%%%%%%%%%%%%%%%%

%%%%%%%%%%%%%%%%%%%%%%%%%%%%%%%%%%%%%%%%%%%%

To obtain these results, we fix $\omega \tau = 1.5$, the number of shots $N_{\text{shots}}^{\text{QNDM}}$ (to determine the precision of the measured detector phase) and the discretization of $\lambda$ in $\Delta \lambda$ (which, eventually, affects the Fourier transform accuracy). 
The \texttt{Qiskit} implementation of the QNDM protocol to extract $\Re(\Glambda)$ and $\Im(\Glambda)$ is provided in Appendix \ref{sec: qiskit_implementation}.

We chose to take $0 \leq \lambda \leq 100$ with discretization of $\Delta \lambda = 0.1$ and $N_{\text{shots}}^{\text{QNDM}} = 10^3$.
The quasi-probability distribution obtained with these parameters is presented in Fig. \ref{fig: QPDV1}a). As it can be seen, for $\Delta = 2$ there is a negative peak, which identifies the presence of quantum effects.

The quasi-probability distribution demonstrates remarkable robustness against variations in both the discretization parameter $\Delta\lambda$ and the number of shots $N_{\text{shots}}^{\text{QNDM}}$ a shown in Fig. \ref{fig: QPDV1}b),
This resilience to discretization stems from the fact that the theoretical QPD, as defined in Eq.~\eqref{eq: P_ND}, is expressed as a sum of delta functions.  
Consequently, the distribution's variance arises solely from statistical noise, which is assumed to have a uniform variance across all frequencies.  
According to Parseval’s identity for the discrete inverse Fourier transform \cite{bracewell1978fourier}, this variance scales linearly with $\Delta \lambda$, following the relation $\sigma_{\mathcal{P}_{ND}} = \Delta\lambda \sigma_{\Glambda}$, where $\sigma_{\mathcal{P}_{ND}}$ and $\sigma_{\Glambda}$ are the variances associated with the noise of the QPD and the quasi-characteristic function, respectively.

An example of this effect is shown in Fig. \ref{fig: QPDV1}b), where the two parameters are chosen to minimize the resources needed to measure $\Glambda$, i.e., $\Delta \lambda = 1$ and $N_{\text{shots}}^{\text{QNDM}}  = 10^2$.
The increase in the discretization with $\Delta \lambda = 1$ leads to a spread of the peaks in the $\Prob_{\text{ND}}(\Delta)$, since it affects the accuracy of the Fourier transform. 
In addition, small spurious features appear around $\pm 3$.
However, the negative regions, i.e., the only feature which we are interested in, are still clearly visible even for $\Delta \lambda = 1$.

To estimate the maximum $\Delta \lambda$ that can be taken, we consider the following argument: in the present model, the statistical error contributes only with fast frequencies at small amplitudes. The highest frequency contribution in Eq.~\eqref{eq: Glambda formal} and \eqref{eq: P_ND} is given by the oscillatory term $\exp \{ \pm i \lambda \Delta_{\max} \}$.  
Since $\Glambda$ oscillates with a maximum period determined by $\Delta_{\max} = \pm3$, according to the Nyquist theorem, the maximum spacing in $\lambda$ that avoids aliasing is  
$\Delta\lambda_{\max} = 2\pi/(2\Delta_{\max}) \approx 1$.
This implies that choosing $\Delta \lambda \approx 1$ is sufficient to capture the relevant oscillatory behavior while avoiding loss of information due to undersampling.

\begin{figure}
    \begin{center}  
    \includegraphics[scale=0.8]{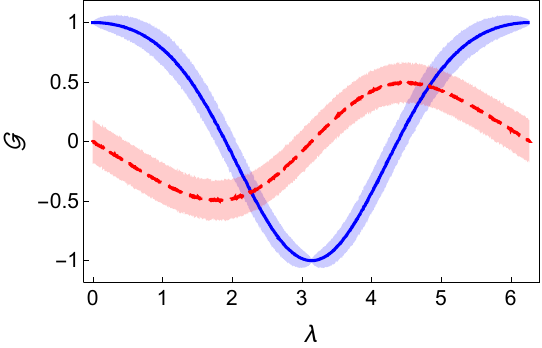}
       \end{center}
    \caption{
        Plot of $\Re(\Glambda)$ (solid blue) and $\Im(\Glambda)$ (dashed red) for $N_{\text{shots}} = 10^3$ and $\Delta\lambda = 0.1$. The solid lines are the average values, while the shaded regions are the statistical errors. }
    \label{fig: Glambda}
\end{figure}

As shown in Fig. \ref{fig: QPDV1}, the negative regions remain stable even for low values of $N_{\text{shots}}^{\text{QNDM}}$, despite the associated increase in statistical error when calculating $\Glambda$.
This can be explained as a feature of the inverse Fourier Transform with the help of Fig. \ref{fig: Glambda}, where $\Re(\Glambda)$ (solid) and $\Im(\Glambda)$ (dashed) are plotted for $N_{\text{shots}}^{\text{QNDM}} = 10^3$ and $\Delta\lambda = 0.1$. Here, for each $\lambda$, we repeated $N=100$ times the calculation of $\Re(\Glambda)$ and $\Im(\Glambda)$ and calculated the mean value and the associated statistical error.
The presence of this statistical noise in $\Glambda$ effectively corresponds to introducing high (and noisy) frequencies.
These are usually relatively small in amplitude compared to the expected ones and, more importantly, they can be eliminated because are outside the interested range of $-\Delta_{max} \leq \Delta \leq \Delta_{max}$.

These two features allow us to reduce both $\Delta \lambda$ and $N_{\text{shots}}^{\text{QNDM}} $ and, therefore, the computational resource but still be able to identify the negative peaks of the quasi-probability density.

%%%%%%%%%%%%%%%%%%%%%%%%%%%%%%%%%%%%%%%%%%%%

%%%%%%%%%%%%%%%%%%%%%%%%%%%%%%%%%%%%%%%%%%%%

\subsection{Resource estimation}\label{sec: resource_estimation}

In this section, we estimate the computational resources required for LG simulations and compare them with those needed for the QNDM method.

For the LG approach, the total resources needed $N_{LG}$ depends on two factors: the number of measurements, $ N_{\text{meas}}^{\text{LG}}=3$, which corresponds to the number of times the circuits must be repeated to compute the LG correlators, and the number of shots, $ N_{\text{shots}}^{\text{LG}} $, that is, the number of repetition of the circuit needed obtain the desired statistical accuracy.

For the QNDM method, we have two parameters with the same meaning: $ N_{\text{meas}}^{\text{QNDM}} = 2 $ (since we extract information about the real and imaginary part of $\Glambda$, see Appendix \ref{sec: qiskit_implementation}) and $N_{\text{shots}}^{\text{QNDM}}$ that sets the statistical accuracy of the results.
In addition, in this case, we need an additional requirement—the discretization of $ \lambda $—which affects the precision of the Fourier Transform. Given the range $ 0 \leq \lambda \leq 100 $, the number of discretization steps, $ N_{\text{steps}}^{\lambda} $, is determined by the step size $ \Delta\lambda $ as $N_{\text{steps}}^{\lambda} =100/\Delta\lambda$.

The total number of circuit evaluations for each method is given by:
\begin{eqnarray}
	 N_{\text{LG}}   &=& N_{\text{meas}}^{\text{LG}} \cdot N_{\text{shots}}^{\text{LG}} \nonumber \\
	N_{\text{QNDM}} &=& N_{\text{meas}}^{\text{QNDM}} \cdot N_{\text{steps}}^{\lambda} \cdot N_{\text{shots}}.
\end{eqnarray}
For the LG simulations, in Fig. \ref{fig: K3_fig}, we need $ N_{\text{LG}} = 3 \cdot 10^3 $ and $ 3 \cdot 10^4 $, respectively.

The robustness of QNDM, discussed in Sec. \ref{sec: QNDM_numerical}, allows us to optimize the parameters without limiting the protocol efficiency.
With an appropriate choice of $N_{\text{steps}}^{\lambda}$ and $N_{\text{shots}}^{\text{QNDM}}$, the computational cost of the QNDM method can be reduced relative to LG simulations, while maintaining the ability to identify negative regions in the QPD. 
For example, for the parameters chosen in Fig. \ref{fig: QPDV1}a), we set $N_{\text{shots}}^{\text{QNDM}} = 10^3$ and $\Delta\lambda = 0.1$, corresponding to $N_{\text{QNDM}} = 2 \cdot 10^6$ circuit evaluations. In contrast, for the parameters used in Fig. \ref{fig: QPDV1}b), we set $N_{\text{shots}}^{\text{QNDM}} = 10^2$ and $\Delta\lambda = 1$, which gives the total number of circuit evaluations to $N_{\text{QNDM}} = 2 \cdot 10^4$. 

Thus, the resources needed in the two approaches are comparable despite the fact that the LGIs often fail to identify the violation of MR.
This makes the QNDM approach both more reliable and efficient in terms of computational or experimental resources.

%%%%%%%%%%%%%%%%%%%%%%%%%%%%%%%%%%%%%%%%%%%%%%%

\begin{figure}
    \begin{center}
        \includegraphics[width=0.8\linewidth]{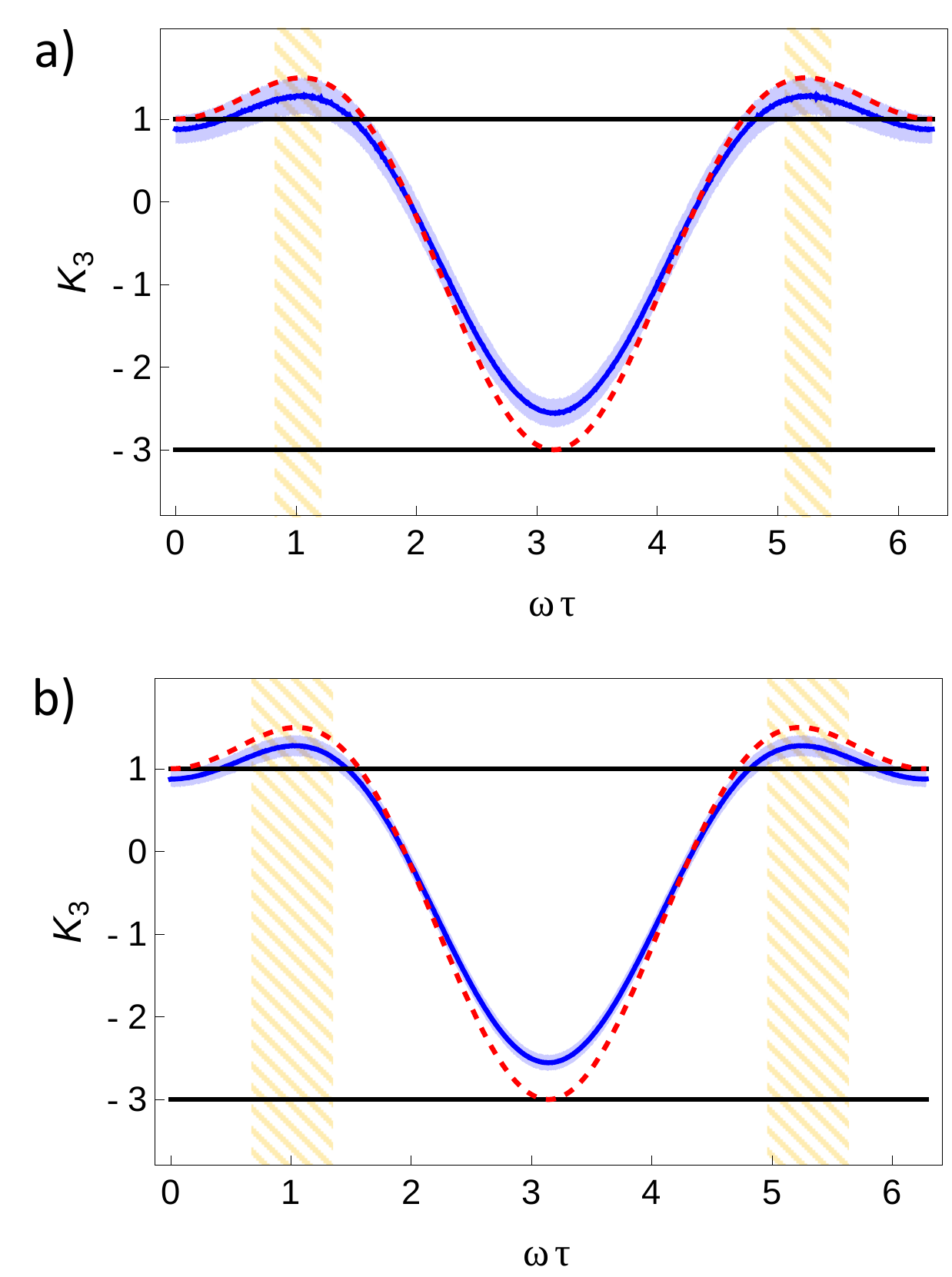}
    \end{center}
    \caption{Plot of LG noise simulations for $ K_3 $ with $ N_{\text{shots}}^{\text{QNDM}} = 10^3 $ shots (a) and $ N_{\text{shots}}^{\text{QNDM}} = 10^4 $ (b). The solid blue line represents the simulated average $ K_3 $ curve, while the dashed red line corresponds to the theoretical prediction. The blue-shaded region indicates the statistical uncertainty. The yellow-shaded region highlights the range of $ \omega \tau $ for which, considering statistical errors, we can confidently assert that the LGIs are violated.}
    \label{fig: K3_fig_noise}
\end{figure}

%%%%%%%%%%%%%%%%%%%%%%%%%%%%%%%%%%%%%%%%%%%%%%%

\section{Realistic simulations with environmental noise}

The results presented in the previous section consider only statistical noise arising from the finite number of experiment repetitions, i.e. the number of shots. However, these results do not account for environmental noise in the quantum device. Since fault-tolerant quantum computers are still at least a decade away, current platforms are classified as noisy intermediate-scale quantum (NISQ) ones. For this reason, assessing the feasibility of implementing the proposed algorithms on today’s quantum hardware requires accounting for environmental noise in our simulations.
This analysis is particularly relevant for the QNDM approach that requires storing the information in the detector phase. Since the quantum phases are particularly sensible to the decoherence and dephasing effects, it might seem that the QNDM approach is not reliable for realistic implementations.

In this section, we extend the analysis of the simulations discussed previously by incorporating a noise model. To perform these simulations, we used IBM's quantum simulators \cite{IBM_docs}, which emulate the behaviour of their Quantum Processing Units (QPUs). These simulators provide a realistic framework for modelling noise, ensuring that the results obtained closely resemble those expected from executions on real quantum hardware. For our simulations, we specifically used the \texttt{FakeBogotaV2} simulator, which replicates the characteristics of a $5-$qubit IBM quantum processor \cite{IBM_docs}.

The results of the noisy simulations of $K_3$ are presented in Fig. \eqref{fig: K3_fig_noise}. As in the noiseless scenario, we took $N_{\text{shots}}^{\text{LG}} = 10^3$ [Fig. \ref{fig: K3_fig_noise}a)] and $N_{\text{shots}}^{\text{LG}}  = 10^4$ [Fig. \ref{fig: K3_fig_noise}b)]. 
As we can see, even the simulated average $K_3$ (solid blue curve) deviates from the expected one (dashed red curve).
More importantly, the regions in which the LGIs are undoubtedly violated are extremely reduced. More quantitatively, for $N_{\text{shots}}^{\text{LG}}  = 10^3$ the region is reduced to the $15\%$ of the total $\omega\tau$ space, while for $N_{\text{shots}}^{\text{LG}}  = 10^4$ this region is reduced to the $24\%$. Indeed, we find that $N_{\text{shots}}^{\text{LG}}  = 10^3$ is a limiting number and a further reduction leads to the vanishing of the LGIs violation interval.

\begin{figure}
    \begin{center}
        \includegraphics[width=0.8\linewidth]{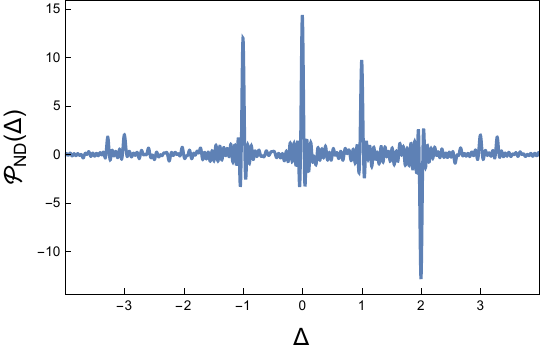}
    \end{center}
    \caption{Plot of the quasi probability $\Prob_{\text{ND}}(\Delta)$ for $\Delta \lambda = 1 $ and $N_{\text{shots}}^{\text{QNDM}}  = 10^{2}$, when both statistical and hardware noise is taken into account.} 
    \label{fig: QPDV1_noise}
\end{figure}

In contrast, the QNDM method, as shown in Fig. \eqref{fig: QPDV1_noise}, exhibits robustness to hardware noise and statistical uncertainty. For this analysis, we chose a set of parameters that minimize the computational resources needed to implement the QNDM circuit, i.e. $\Delta \lambda = 1$ and $N_{\text{shots}}^{\text{QNDM}}  = 10^2$. In this case, the negative regions associated with the violation of MR are still clearly visible. 

The resource estimation computation follows the same procedure as described in Sec. \ref{sec: resource_estimation}. For the parameters chosen in Fig. \eqref{fig: QPDV1_noise}, we set $N_{\text{shots}}^{\text{QNDM}}  = 10^2$ and $\Delta\lambda = 1$. This results in a total of $2 \cdot 10^4$ circuit evaluations. In contrast, the number of shots required to identify violations of LGIs are at least $N_{\text{shots}}^{\text{LG}}  \gtrsim 10^{4}$, so that the resource requirements for the LG approach are of order $3 \cdot 10^4$. Thus, the QNDM method is slightly more efficient in the presence of noise, requiring fewer or comparable resources to identify quantum features in the system.

%%%%%%%%%%%%%%%%%%%%%%%%%%%%%%%%%%%%%%%%%%%%%%%

\section{Conclusions}
\label{sec: conclusions}

We have performed a detailed comparison between two approaches used to identify the violation of the MR condition.
The first one exploits the LGIs and, experimentally, is reduced to measuring the correlators of sequential quantum observables.
The second one, i.e. the QNDM approach, uses a quantum detector to store the desired information about the sequential measurements.
It leads to a quasi-distribution probability whose negative regions unequivocally identify the interference and the violation of the MR condition \cite{solinas2024}.

Using an IBM quantum computer to implement the model and perform the simulations, we were able to account for the finite number of quantum circuit executions, i.e. the experiments, and the corresponding statistical error.
We found that the resources needed, i.e. the number of quantum circuit runs, are of the same order of magnitude (with slight advantages of the QNDM approach).
However, if we take into account that the QPD obtained with the QNDM approach allows us to always identify the violation of MR condition \cite{solinas2024} while the LGI often fails to do it.

These results are confirmed also in the presence of environmental noise which has been simulated using the built-in noise model of the IBM quantum computer.
In this case, the advantage of the QNDM versus LGIs approach is even more surprising if we consider the fact that the quantum phase used to store the information in the first one is usually extremely sensible to dephasing and decoherence.

The root of these results and the robustness of the QNDM approach to both statistical and environmental errors lies in the Fourier transform of the data needed to obtain the QPD.
Both errors introduce high-frequency fluctuations in the quasi-characteristic function obtained through the phase measurement. These are easily smoothed and eliminated by the Fourier transform of the data.
As a consequence, the negative regions of the QPD are barely affected by the presence of the noise.

These results show that the QNDM approach offers a valid alternative to the LGIs.
Not only do they always identify the violation of MR (contrarily to the LGIs) but they also need fewer experimental resources.

%%%%%%%%%%%%%%%%%%%%%%%%%%%%%%%%%%%%%%%%%%%%%%%

\section*{Acknowledgements}
The authors acknowledge financial support from INFN. 

%%%%%%%%%%%%%%%%%%%%%%%%%%%%%%%%%%%%%%%%%%%%%%%

%%%%%%%%%%%%%%%%%%%%%%%%%%%%%%%%%%%%%%%%%%%%%%%

\appendix

\begin{appendices}

%%%%%%%%%%%%%%%%%%%%%%%%%%%%%%%%%%%%%%%%%%%%%%%%

\section{\texttt{Qiskit} implementation}\label{sec: qiskit_implementation}

In this section, we present the \texttt{Qiskit} implementation of the gates and circuits with which we performed the simulations. We first present the implementation of the initial state, followed by the evolution operator and the interaction operator. Finally, we outline the implementation of the QNDM protocol.

\begin{figure}     
    \includegraphics[width=0.4\textwidth]{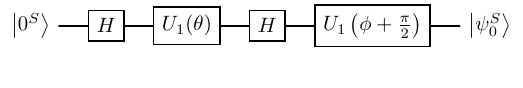}     
    \caption[Initial state implementation]{Initial state implementation. }     
    \label{fig: initial state implementation} 
\end{figure}

The generic one-qubit state can be parametrized by $\theta$ and $\phi$ and is given by: 
\begin{equation} 
    \ket{\psi^S_0} = \cos\frac{\theta}{2} \ket{0} + e^{i\phi} \sin\frac{\theta}{2} \ket{1}.
\end{equation} 
The set of gates needed to perform the operation:
\begin{equation} 
    \ket{0} \longrightarrow \cos\frac{\theta}{2} \ket{0} + e^{i\phi} \sin\frac{\theta}{2} \ket{1} 
\end{equation} 
is given by: 
\begin{equation}\label{eq: initial state gates} 
    U_{\text{init}}(\theta, \phi) = U_{1}\qty(\phi + \frac{\pi}{2}) H U_{1}\qty(\theta) H, 
\end{equation} 
where: 
\begin{equation} U_{1}\qty(\alpha) = 
    \begin{pmatrix}       1 & 0 \\
     0 & e^{i\alpha}     
    \end{pmatrix}. 
\end{equation} 
Fig. \eqref{fig: initial state implementation} shows the circuital implementation for the initial state.\ The initial state for our simulations was $\ket{\psi_{0}^{S}} = (\ket{0} + i \ket{1})/\sqrt{2}$, and can be obtained by \eqref{eq: initial state gates} by choosing $\theta = \pi/2$ and $\phi = \pi/2$. 

The unitary evolution $U  = e^{-i\alpha \hat{\sigma}_{x}}$ corresponding to the rotation generated by the Pauli matrix $\hat{\sigma}_{x}$ is given by: 
\begin{equation} 
    U(\alpha) = e^{-i\alpha \hat{\sigma}{x}} = 
    \begin{pmatrix} 
        \cos\alpha & -i\sin\alpha \\ 
        -i\sin\alpha & \cos\alpha 
    \end{pmatrix} . 
\end{equation} 
This operation can be decomposed in terms of quantum gates as: 
\begin{equation} 
    e^{-i\alpha \hat{\sigma}{x}} = H \cdot R_{z}(2\alpha ) \cdot H, 
\end{equation} 
where $H$ is the Hadamard gate, and $R_{z}(\theta)$ represents a phase rotation about the $z$-axis, defined as: 
\begin{equation} 
    R_{z}(\theta) 
    = e^{-i\frac{\theta}{2} \hat{\sigma}_{z}} = 
    \begin{pmatrix} 
        e^{-i\frac{\theta}{2}} & 0 \\ 0 & e^{i\frac{\theta}{2} } 
    \end{pmatrix}. 
\end{equation}
Thus, the decomposition shows that a rotation about the $x$-axis can be achieved using Hadamard transformations and a single rotation about the $z$-axis.

The interaction operator between system and detector qubits is $\hat{u}(\lambda) = \exp\qty{i \lambda \hat{\sigma}{z}^{S} \otimes \hat{\sigma}{z}^{D}}$. In terms of the basis $\qty{\ket{0_S 0_D}, \ket{0_S 1_D}, \ket{1_S 0_D}, \ket{1_S 1_D}}$, the matrix form of $\hat{u}$: 
\begin{equation} 
    \hat{u}(\lambda) =  e^{i \lambda \hat{\sigma}{z}^{S} \otimes \hat{\sigma}{z}^{D}} = 
    \begin{pmatrix} 
        e^{i\lambda} & 0 & 0 & 0 \\         
        0 & e^{-i\lambda} & 0 & 0 \\         
        0 & 0 & e^{-i\lambda} & 0 \\         
        0 & 0 & 0 & e^{i\lambda}     
    \end{pmatrix} 
\end{equation} 

The \texttt{Qiskit} implementation is given by (a part a $e^{i\lambda}$ factor): 
\begin{equation} 
    \hat{u}(\lambda) = CX_{1,2} \qty[\mathbb{I} \otimes U_{1}(-2\lambda)]  CX_{1,2} 
\end{equation} 
where we denoted $1$ as the system qubit, and with $2$ the detector one.

\begin{figure*}[t!]
        \centering
        \includegraphics[width=0.8\textwidth]{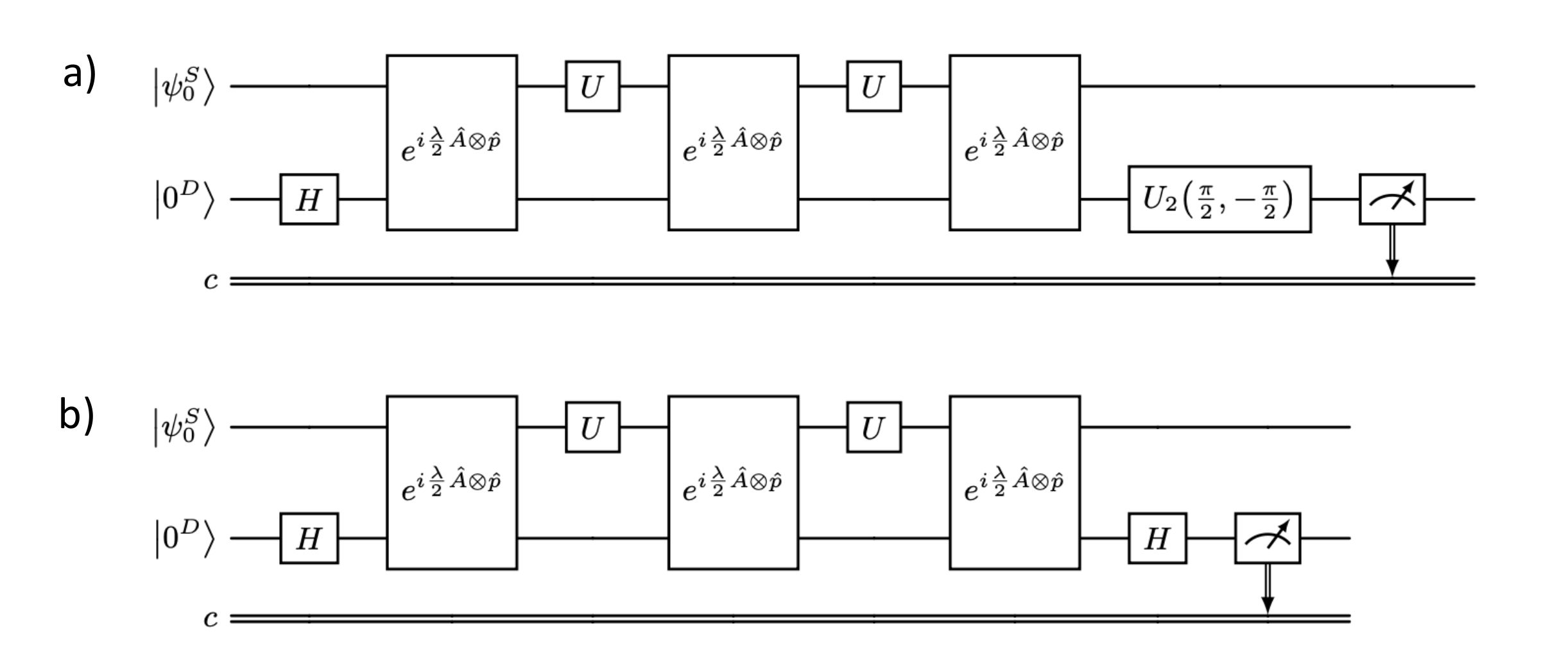}
        \caption[QNDM circuit to extract $\Im{G}$]{QNDM circuit to extract $\Im\qty(\Glambda)$ (a) and $\Re\qty(\Glambda)$ (b). Here, $\ket{\psi_{0}^{S}}$ is the initial state of the system qubit as in Fig. \eqref{fig: initial state implementation}, and $c$ corresponds to the classical register where the measurement outcomes are stored. The gates $U_2$ and $H$ are explained in the text and represent a standard method to extract the real and imaginary parts of the accumulated phase in the detector qubit.}
        \label{fig: QNDM_circuit}
\end{figure*}

In the QNDM protocol, two distinct quantum circuits are required for each $ \lambda $ to extract $ \Glambda $. Specifically, one circuit is used to extract $ \text{Re}(\Glambda) $ and another for $ \text{Im}(\Glambda) $, as illustrated in Fig. \ref{fig: QNDM_circuit}a) and \ref{fig: QNDM_circuit}b), respectively. The output of the circuits represents the probabilities of occurrence of the detector state $ \ket{0^D} $ and $ \ket{1^D} $, which are functions of the coupling parameter $ \lambda $.

The real and imaginary part of $\Glambda$ can be directly measured with standard interferometric techniques \cite{solinas2021, Solinas_2023}. More specifically, to extract the real part of $ \Glambda $, we rotate the basis with an Hadamard gate $ H $, while for the imaginary part, we use the $ U_2 $ gate, which is defined as:
\begin{equation}
 U_{2}(\alpha, \beta) = \frac{1}{\sqrt{2}} \begin{pmatrix} 1 & -e^{i\beta} \\ e^{i\alpha} & e^{i(\alpha+\beta)}
    \end{pmatrix}.
\end{equation}
In our specific case, the rotation parameters are $ \alpha = \pi/2 $ and $ \beta = -\pi/2 $, so that:
\begin{equation}
 U_{2}\qty(\frac{\pi}{2}, -\frac{\pi}{2}) = \frac{1}{\sqrt{2}}\begin{pmatrix}
    1 & i \\ i & 1
    \end{pmatrix}.
\end{equation}
The circuital implementation is provided in Fig. \ref{fig: QNDM_circuit}.

\section{Note on the complementarity of LGIs}

In our simulations, we chose the standard set-up, where the interaction operator and correlator are perpendicular. In this case, the LGIs are complementary in the sense that, for every $\omega\tau\in [0,2\pi)$, there is a $K_i$ violating the classical limits. Specifically, the LGI \eqref{eq: LGI_3} is violated for $0 \leq \omega\tau \leq \pi/2$ and $3\pi/2 \leq \omega\tau \leq 2\pi$, and satisfied in the range $\pi/2 \leq \omega\tau \leq 3\pi/2$. Conversely, the LGI associated with $K_2$ is violated in the interval $\pi/2 \leq \omega\tau \leq 3\pi/2$, while satisfied for $0 \leq \omega\tau \leq \pi/2$ and $3\pi/2 \leq \omega\tau \leq 2\pi$. Lastly, LGI for $K_1$ is always satisfied. Notably, this complementarity among the LGIs is not guaranteed in general \cite{halliwell21,huelga95}. In practice, it is usually necessary to study higher orders of the LGIs \cite{Emary_2014}, called $n$-measurement Leggett-Garg strings:
\begin{align}
 K_n = C_{21} + C_{32} + C_{43} + \dots + C_{n(n-1)} - C_{n1}.
\end{align}
This procedure increases the resource requirements for the simulation. In comparison, the proposed QNDM protocol detects violations of macrorealism with a fixed number of detector interactions, independent of the specific system being studied.

\end{appendices}

\end{document}